 \documentstyle[aps]{revtex}  
\newcommand{\tsub}[1]{_{\mbox{\scriptsize#1}}}

\newcommand{\tfrac}[2]{\mbox{$\small\frac{#1}{#2}$}}
\newcommand{\quarterthin}{\kern 0.0417em}
\newcommand{\comm}[2]{[ \quarterthin #1 , #2 \quarterthin ]}
\newcommand{\bra}[1]{\langle#1|}
\newcommand{\ket}[1]{|#1\rangle}
\newcommand{\ev}[1]{\langle#1\rangle}
\newcommand{\mel}[3]{\bra{#1}#2\ket{#3}}
\newcommand{\thin}{\thinspace}

\begin{document}
\draft


 \twocolumn[\hsize\textwidth\columnwidth\hsize  
 \csname @twocolumnfalse\endcsname              

\title{An SU(4) Model of High-Temperature Superconductivity and
Antiferromagnetism}

\author{Mike Guidry$^{(1)}$, Lian-Ao Wu$^{(2)}$, 
Yang Sun$^{(1)}$ and Cheng-Li Wu$^{(3,4)}$,
}
\address{
$^{(1)}$Department of Physics and Astronomy, University of Tennessee,
Knoxville, Tennessee 37996\\
$^{(2)}$Department of Physics, Jilin University, Changchun, Jilin 130023, PRC\\
$^{(3)}$Department of Physics, Chung-Yuan Christian University, 
Chungli, Taiwan 320, ROC\\
$^{(4)}$National Center for Theoretical Science, Hsinchu, Taiwan 300, ROC}

\date{\today}
\maketitle

\begin{abstract}
We present an $SU(4)$ model of high-temperature superconductivity having many
similarities to dynamical symmetries known to play an important role in microscopic
nuclear structure physics and in elementary particle physics.  Analytical
solutions in three dynamical symmetry limits of this model are found: an
$SO(4)$ limit associated with antiferromagnetic order; an $SU(2) \times SO(3)$
limit that may be interpreted as a $d$-wave pairing condensate; and an $SO(5)$
limit that may be interpreted as a doorway state between the antiferromagnetic
order and the superconducting order.  The model suggests a phase diagram in
qualitative agreement with that observed in the cuprate superconductors. The
relationship between the present model and the $SO(5)$ unification of
superconductivity and antiferromagnetic order proposed by Zhang is discussed.
\end{abstract}

\pacs{}

 ]  

\narrowtext

There are compelling arguments that the  mechanism leading to
high-temperature superconductivity 
does not correspond to ordinary BCS $s$-wave pairing.  Although experimental
evidence implicates singlet (hole) pairs as the carriers of the supercurrent,
the interaction leading to the formation of the
singlet pairs appears not to be
the traditional lattice phonon mechanism underlying the
BCS theory, but rather seems to be a collective electronic interaction. 
Furthermore,
the pairing gap is anisotropic, with nodes in the $k_x$--$k_y$ 
plane strongly 
suggestive of $d$-wave hybridization in the 2-particle
wavefunctions, and the mechanism responsible for superconductivity in the
cuprates is thought to be closely related to the unusual
antiferromagnetic (AF) insulator properties of
their normal states. 
 
Contrary to the case for BCS
superconductors, the formation of Cooper pairs and the formation
of a superconducting (SC) condensate of those pairs
in high-$T\tsub c$ compounds may be distinct, with
pair formation corresponding to a higher temperature scale
than the condensation
of the pairs into the SC state.  That is, there appear to be at
least two distinct energy scales associated with the formation of the
high-temperature SC state.  This is reminiscent
of grand unified theories in elementary particle
physics, where qualitatively different physical phases result from a hierarchy
of symmetry breakings occurring on different energy (temperature) scales.  This
finds its most natural explanation in a Lie group structure that is broken
spontaneously (and perhaps explicitly) down to subgroups at different
characteristic energy scales.

\section{Dynamics and Symmetries}
 
Such observations argue strongly for a theory based on continuous symmetries of
the dynamical system that is capable of
describing more sophisticated pairing than found in the simple BCS picture
(which is
described by a single complex order parameter), and capable of unifying
different collective modes and phases on a equivalent footing.  Then such
fundamentally different physics as antiferromagnetic order and
superconductivity can 
emerge from the same effective Hamiltonian as concentration
variables (e.g., doping parameters) are varied.

\subsection{Fermion Dynamical Symmetries}

For approximately the same period of time 
that the high-$T\tsub c$ compounds have
been known, techniques based on dynamical symmetries in fermion degrees of
freedom that are capable of satisfying the preceding conditions have been in
development in the field of nuclear
structure physics.  There it has proven fruitful to ask the following
questions:  what are the most important collective degrees of freedom in the
low-lying spectrum of complex nuclei, what are the microscopic
many-body quantum operators
that create and annihilate these modes, and what is the commutator algebra
obeyed by this set of operators.  

Systematic investigation of these questions
has led to strong confirmation of the following set of conjectures about the
nuclear many-body system:  (1)  Low-lying collective modes are in approximate
one-to-one correspondence with dynamical symmetries in the fermion degrees of
freedom.  (A system possesses a dynamical symmetry if it has a
Hamiltonian that can be expressed as a polynomial in
the Casimir invariants of a subgroup
chain.) (2) A dynamical symmetry associated with low-lying collective
modes is associated with a Lie algebra and its subalgebras that are
formed from a set
of fermion operators closed under commutation.  (3)  Different dynamical
symmetry subgroup chains arising from the same highest symmetry group  
are associated with fundamentally different phases of the theory. 
These dynamical symmetries are characterized by different
collective modes and the corresponding
 phases are unified in the highest group, just as
grand unified theories are unified in the higher groups of the symmetry
breaking chain.  (4) The unification implied by the preceding point suggests 
that the many low-lying collective states formed by systematic filling of
valence shells in heavy nuclei are in reality different projections in an
abstract multidimensional space of the same state. Equivalently, the 
different states are transformed into each other by the generators of the
symmetry.  Thus, the systematics of collective modes and phase transitions as
a function of concentration variables are specified by the group structure.

It has been demonstrated that dynamical symmetries of the type
described in Ref.\ \cite{wu94} are realized to remarkably high accuracy in the
spectrum and the wavefunctions of large-scale numerical calculations
using the Projected Shell Model
\cite{sun98}.  Since this model is known to
give extremely good agreement with a broad range of experimental data
(see, e.g., Ref.\ \cite{sun97}), this provides rather conclusive proof that
these dynamical symmetries are strongly realized in the low-lying states
of complex nuclei.  This raises the issue of whether similar symmetries might
be found in other complex many-body fermion systems such as those important
in condensed matter.  We may expect that the nuclear and condensed matter
systems have many similarities that could make this consideration fruitful:
both are composed of interacting
fermions and both are ultimately many-body systems 
that are only approximately describable by mean-field ideas.

\subsection{The Zhang SO(5) Model}

S. C. Zhang and collaborators have introduced ideas bearing many
similarities to these into the high temperature superconductor discussion
\cite{zha97}.
Motivated by a desire to unify AF and SC order
parameters, Zhang et al have assembled these into a 5-dimensional vector order
parameter, and have then constructed an $SO(5)$ group that rotates the
AF order parameters into the SC ones.  This
construction is based explicitly on the assumption of $d$-wave pairing in the
SC state.  

In this paper we proceed differently.  We start, not from the desire to unify
two particular phases, but from identification of a closed algebra associated
with a general set of fermion pairing and particle--hole operators defined on
a periodic lattice.  Nevertheless, we shall find that we recover Zhang's
$SO(5)$ symmetry as a subgroup of a more general $U(4)$ symmetry if 
certain commutators of the full $SU(4)$ algebra are set to zero.  Thus,
much of the extensive recent discussion of the Zhang $SO(5)$ symmetry applies
directly to the results of this paper.  

However, the present paper extends this
discussion substantially:  (1) 
The $SO(5)$ subgroup is embedded in a larger algebra defined microscopically in
the fermion degrees of freedom,
which implies constraints on the $SO(5)$ subgroup. 
(2) The $SU(4)$ highest symmetry
has subgroups  in addition to $SO(5)$ that may be relevant to the condensed
matter problem in general and the cuprate superconductors in particular. 
(3) 
We frame the discussion, not in terms of an approximate symmetry of
a Hubbard or $t$--$J$ Hamiltonian, but in terms of an exact {\em dynamical
symmetry} constructed using the Casimir invariants of group chains.   
We shall present a more detailed discussion of the relationship between the
present model and the Zhang $SO(5)$ model.

\section{Dynamical Symmetry Method}

The dynamical symmetry method applied here corresponds schematically to
the following algorithm:

0.  Assume the following conjecture:  {\em All strongly collective
modes in fermion (or boson) many-body systems can be put
into correspondence with a closed algebra defining a dynamical symmetry of
the sort described below.}  This is a conjecture, but  
there is so much evidence in support of it from various fields of physics
that it is almost a theorem:
Strongly correlated motion implies a symmetry of the dynamics described by
a Lie algebra in the second-quantized operators implementing that motion.

1.  Identify, within a suitable ``valence space'', degrees of
freedom that one believes are physically relevant for the problem at hand,
guided by phenomenology, theory, and general principles.  In the present
case, that reduces to defining a minimal set of operators that might
be important to describe superconductivity and antiferromagnetism
on a spin lattice.

2.  Try to close a commutation algebra (of manageable dimension) with the
second-quantized operators creating and annihilating the modes chosen in step
1.  If necessary, approximate these operators, or add additional ones to
the set if the algebra does not close naturally. In the present context,
the simple $g(k)$ form-factor introduced below
is an example of simplifying things to close the
algebra.

3.  Use standard Lie algebra theory to identify 
relevant subalgebra
chains that end in algebras for conservation laws that one
expects to be obeyed for the problem at hand.  
In the present example, we require all group chains
to end in $U(1) \times SU(2)$, corresponding to 
an algebra implementing conservation of charge and spin.

4.  Construct dynamical symmetry Hamiltonians (Hamiltonian that are
polynomials in the Casimir invariants of a group chain) for each chain.
Each such group chain thus defines a wavefunction basis labeled  by the
eigenvalues of chain invariants (the Casimirs and the elements of the
Cartan subalgebras), and a Hamiltonian that is diagonal in that basis
(since it is constructed explicitly from invariants).  Thus, the
Schoedinger equation is solved analytically for each chain, by
construction.

5.  Calculate the physical implications of each of these dynamical
symmetries by considering the wavefunctions, spectra, and transitional
matrix elements of physical relevance.  This is tractable, because the
eigenvalues and eigenvectors were obtained in step 4. Consistency of the
symmetry requires that transition operators be related to group
generators; otherwise transitions would mix irreducible
multiplets and break the symmetry.

6.  If step 5 suggests that one is on the right track (meaning that a wise
choice was made in step 1), one can write the most general Hamiltonian for the
system in the model space, which is just a linear combination of all the
Hamiltonians for the symmetry group chains.
Since the Casimir operators of different group chains do not generally commute
with each other, a Casimir invariant for one group chain may be a
symmetry-breaking term for another group chain.  Thus the competition between
different dynamical symmetries and the corresponding phase transitions can be
studied.

7.  The symmetry-limit solutions may be used as a starting
point for more ambitious calculations that incorporate symmetry
breaking.  Although no longer generally analytical, such more realistic
approximations may be solved
by perturbation theory around the symmetry solutions
(which are generally non-perturbative, so this is perturbation theory
around a non-perturbative minimum), 
by numerical diagonalization of symmetry breaking terms, or by 
coherent state or other mean-field approximations.

Representative application of these ideas for both fermion and
boson systems may be found in  nuclear \cite{wu94,IBM}, particle \cite{Bi94},
molecular \cite{Ia95}, and polymer physics \cite{Ia99}.  We also note that the
general idea of symmetry having dynamical implications lies at the heart of
local gauge field theories in particle physics, though the details and
methodologies in that case differ from the ones used here.
 
The only approximation in the dynamical symmetry approach outlined above is the
space truncation.  If all degrees of freedom are incorporated, 
this defines an exact
microscopic theory.  Of course, in practical calculations only a few carefully
selected degrees of freedom can be included and the effect of the excluded
space must be incorporated by renormalized interactions in the truncated
space.  It follows that 
the validity of such an approach hinges on a wise choice of the
collective degrees of freedom and sufficient phenomenological or theoretical
information to specify the corresponding effective interactions of the
truncated space.

\section{The SU(4) Model}

Let us now introduce a
mathematical formalism that provides a
systematic implementation of the dynamical symmetry
procedure for a particular physically-motivated choice of
operators.

\subsection{Choice of Operators}

The success of the dynamical symmetry method depends on a wise selection of the
operators that describe the low-energy degrees of freedom for
the system.  In the case of cuprate superconductors, we know that (unlike for
normal superconductors) antiferromagnetism and superconductivity lie very near
each other in the phase diagram.  Further, data suggest that the 
SC phases are associated with Cooper pairs of spin-singlet 
electron holes having $d$-wave geometry.  Finally, we expect that the physical
system must conserve both charge and spin.  These observations suggest that at
a minimum we need $d$-wave singlet pairs and operators associated with
antiferromagnetism, spin operators, and charge operators entering the theory on
an equivalent footing.  Let us now construct a minimal closed algebra 
that incorporates these degrees of freedom
\cite{minimalnote}. 

\subsection{The Algebra}

We begin by defining the following lattice fermion operators:
\begin{eqnarray}
p_{12}^\dagger&=&\sum_k g(k) c_{k\uparrow}^\dagger
c_{-k\downarrow}^\dagger
\qquad p_{12}=\sum_k g^*(k) c_{-k\downarrow} c_{k\uparrow} \nonumber
\\ q_{ij}^\dagger &=& \sum_k g(k) c_{k+Q,i}^\dagger c_{-k,j}^\dagger
\qquad q_{ij} = (q_{ij}^\dagger)^\dagger
\label{E1}
\\ Q_{ij} &=& \sum_k c_{k+Q,i}^\dagger c_{k,j} \qquad S_{ij} = \sum_k
c_{k,i}^\dagger c_{k,j} - \tfrac12 \Omega \delta_{ij}  \nonumber
\end{eqnarray}
where $c_{k,i}^\dagger$ creates a fermion of momentum $k$ and
spin projection $i,j= 1 {\rm\ or\ }2 = \ \uparrow$ or
$\downarrow$, $Q=(\pi,\pi,\pi)$ is an AF ordering vector,
$\Omega$ is the lattice degeneracy, and we approximate a $d$-wave form-factor by
$$
 g(k) = {\rm sign} (\cos k_x -\cos k_y) = \pm 1
$$ 
with $g(k+Q) = -g(k)$ and
$\left| g(k) \right| = 1$ (see Refs.\ \cite{hen98,rab98}).
Using the usual fermion anticommutators, 
we deduce the following commutation relations among the operators
of Eq.\ (\ref{E1}):
\begin{eqnarray}
\comm{p_{12}\hspace{1pt}}{\hspace{1pt}p_{12}^{\dagger}}&=&-S_{11}-S_{22}
\nonumber\\
\comm{q_{ij}\hspace{4pt}}{\hspace{4pt}q_{kl}^\dagger} &=& -\delta_{ik}S_{lj} 
-\delta_{il}S_{kj} -
\delta_{kj}S_{li} - \delta_{jl}S_{ki}
\nonumber\\
\comm{p_{ij}\hspace{2pt}}{\hspace{4pt}q_{kl}^\dagger} &=& \delta_{ik}Q_{lj}
+\delta_{il}Q_{kj} -
\delta_{kj}Q_{li} - \delta_{jl}Q_{ki}
\nonumber\\
\comm{S_{ij}\hspace{2pt}}{\hspace{2pt}S_{kl}}&=&\delta_{jk}S_{il} -
\delta_{il}S_{kj}
\nonumber\\
\comm{Q_{ij}}{Q_{kl}}&=&\delta_{jk}S_{il} - \delta_{il}S_{kj}
\\
\comm{S_{ij}\hspace{2pt}}{\hspace{2pt}p_{kl}^\dagger}&=&\delta_{ik}p_{kl}^\dagger
- 
\delta_{jl}p_{ik}^\dagger
\nonumber\\
\comm{S_{ij}\hspace{2pt}}{\hspace{4pt}q_{kl}^{\dagger}}&=&\delta_{ik}q_{il}^{\dagger}
+
\delta_{jl}q_{ik}^\dagger
\nonumber\\
\comm{Q_{ij}}{\hspace{2pt}p_{kl}^\dagger}&=&\delta_{jk}q_{il}^\dagger - 
\delta_{jl}q_{ik}^\dagger
\nonumber\\
\comm{Q_{ij}}{\hspace{4pt}q_{kl}^\dagger}&=&\delta_{jk}p_{il}^\dagger +
\delta_{jl}p_{ik}^\dagger
\nonumber
\end{eqnarray}
Thus, this set of 16 
operators is closed under commutation and generates a Lie
algebra.  Detailed examination indicates that the algebra is
 associated with the group $U(4)$ and has the subgroup chains  
\begin{eqnarray}
&\supset& SO(4) \times U(1) \supset SU(2)\tsub{s} \times U(1) \nonumber \\ U(4)
\supset SU(4) &\supset& SO(5) \supset SU(2)\tsub{s} \times U(1)
\label{eq3}
\\ &\supset& SU(2)\tsub{p}
\times SU(2)\tsub{s} \supset SU(2)\tsub{s} \times U(1) \nonumber
\end{eqnarray}
that end in the subgroup $SU(2)\tsub{s} \times U(1)$ representing spin and
charge conservation.  
The physical interpretation is aided by
rewriting the generators of the $U(4)$ algebra as
\begin{eqnarray}
Q_+&=&Q_{11}+Q_{22} = \sum_k (c_{k+Q\uparrow}^\dagger
c_{k\uparrow} + c_{k+Q\downarrow}^\dagger c_{k\downarrow}) \nonumber
\\
\vec S &=& \left( \frac{S_{12}+S_{21}}{2},
                \ -i \, \frac {S_{12}-S_{21}}{2},
                \ \frac {S_{11}-S_{22}}{2} \right) \nonumber
\\
\vec {\cal Q} &=& \left(\frac{Q_{12}+Q_{21}}{2},\ -i\, \frac{Q_{12}-Q_{21}}{2},
\ \frac{Q_{11}-Q_{22}}{2} \right)
\label{E4} 
\\
\vec \pi^\dagger &=& \left(
\tfrac i2\, (q_{11}^\dagger - q_{22}^\dagger), \
\tfrac 12 (q_{11}^\dagger + q_{22}^\dagger),
\ -\tfrac i2\, (q_{12}^\dagger + q_{21}^\dagger) \right) \nonumber
\\
\vec \pi&=&(\vec \pi^\dagger)^\dagger
\quad D^\dagger = p^\dagger_{12}
\quad D = p_{12}
\quad M=\tfrac12 (n-\Omega) \nonumber
\end{eqnarray}
where $Q_+$ generates charge density waves,
$\vec S$ is the spin operator, $\vec {\cal Q}$ is the staggered magnetization,
and $\vec \pi^\dagger,
\vec \pi$ create and annihilate
 triplet $d$-wave pairs (see Ref.\  \cite{zha97}), the operators
$D^\dagger$ and $D$ are associated with singlet $d$-wave pairs,
$n$ is the electron number operator,
and $M$ is the charge operator.
Properties of this group structure are summarized in Tables I and II,
 and Fig.\ 1.

Notice in this context that we require exact conservation of charge and spin
for our dynamical symmetry solutions
because we have not introduced approximations that
violate these symmetries.  Although it is common to refer to superconductivity
as resulting from violation of particle number, this is a statement about an
approximate solution.  In the exact solution and in nature, particle number is
conserved \cite{spontaneousnote}.  
In a later section we shall introduce
useful {\em approximate solutions}
through coherent state methods that lead to spontaneous
symmetry breaking and to intrinsic states violating particle number, but our
dynamical symmetry solutions conserve charge and spin exactly.

\subsection{The Collective Subspace}

The group $SU(4)$  has a quadratic Casimir operator
\begin{equation}
C_{su(4)}=\vec \pi^\dagger \hspace{-2pt}\cdot
\hspace{-2pt}\vec \pi + D^\dagger D +
\vec S \hspace{-2pt}\cdot\hspace{-2pt} \vec S + \vec {\cal Q}
\hspace{-2pt}\cdot\hspace{-2pt} \vec {\cal Q} + M(M-4)
\label{csu4}
\end{equation}
The group is rank-3 and the irreducible representations
(irreps) may be labeled by 3 weight-space
quantum numbers,
$(\sigma_1,\sigma_2,\sigma_3)$.
We assume for the simplest implementation of the model a collective
$d$-wave pair subspace spanned by the following vectors:
\begin{equation}
\ket{S} = \ket{n_x n_y n_z n_s} = 
(\pi_x^\dagger)^{n_x}
(\pi_y^\dagger)^{n_y}
(\pi_z^\dagger)^{n_z}
(D^\dagger)^{n_s}
\ket{0}
\label{collsubspace}
\end{equation}
This collective subspace is associated with irreps of the form
\begin{equation}
(\sigma_1,\sigma_2,\sigma_3) = (\tfrac \Omega2,0,0)
\end{equation}
where $\Omega$ is the number of lattice sites.
The corresponding expectation value of the  $SU(4)$ Casimir
evaluated in these irreps is a constant,
\begin{equation}
\ev{C_{su(4)}}=\tfrac\Omega2(\tfrac\Omega2 + 4)
\end{equation}

The operator $Q_+$ defined in Eq. (\ref{E4})  
is the generator of the $U(1)$ factor in $U(4) \supset U(1) \times SU(4)$.
Physically, it is associated
with charge density
wave excitations in the system.
We note that $Q_+$ commutes with all generators so it annihilates
the state $\ket S$,
\begin{equation}
Q_+\ket{S} = 0 \qquad \mel S{Q_+}S = 0
\end{equation}
Thus, the collective subspace considered in isolation
is associated with an eigenvalue $Q_+ = 0$.
Physically, this
corresponds to exclusion of charge-density wave excitations
in the low-lying collective subspace
of the effective theory \cite{cdwavenote}.

The dimensionality of the full space is $2^{2\Omega}$.  The dimensionality of
the collective subspace is much smaller, scaling approximately as 
$\Omega^4$:
\begin{equation}
{\rm Dim\thin} (\tfrac{\Omega}{2},0,0) = \tfrac{1}{12}
(\tfrac\Omega 2 + 1)(\tfrac \Omega 2 + 2)^2
(\tfrac \Omega 2 + 3)
\end{equation}
Thus for small lattices it
is possible to enumerate all states of the collective subspace 
in a simple
model where observables can be calculated analytically.

\subsection{SU(4) Model Hamiltonian}

The most general 2-body Hamiltonian within the
$d$-wave pair space consists of a linear combination of (quadratic) Casimir
operators $C_g$ for all subgroups $g$ \cite{threebodynote}
$$
H =H_0+\sum_{g}H_{g}C_{g},
$$
where $H_0$ and $H_{g}$ are parameters and  the Casimir
operators $C_{g}$ are (see Table I)
\begin{eqnarray}
C_{SO(5)} &=&\vec \pi^\dagger \cdot \vec \pi + \vec S \cdot \vec
S +M(M-3) \nonumber \\
C_{SO(4)} &=& \vec {\cal Q} \cdot \vec {\cal Q} + \vec S
\cdot\vec S \nonumber \\
C_{SU(2)\tsub{p}} &=&D^\dagger D +M(M-1) 
\label{casimirs}
\\
C_{SU(2)\tsub s} &=& \vec S \cdot \vec S
\nonumber \\
C_{U(1)} &=&M \mbox{ and }M^2.
\nonumber
\end{eqnarray}
For fixed electron
number the terms in $M$ and
$M^2$ in Eq.\ (\ref{casimirs}) are constant.
The term $H_0$ is a quadratic function of particle number
 and may be parameterized as
$$
H_0 =\varepsilon  n + \tfrac12 \mbox{v}n(n-1),
$$
where $\varepsilon$ and $\mbox{v}$ are
the effective  single-electron energy and the average
two-body interaction in zero-order approximation, respectively.
Thus the Hamiltonian can be written as
\begin{eqnarray} H&=&H_0 + V = \varepsilon n -\tfrac12 \mbox{v} n(n-1)+V
\label{ham}
\\
V&=&
-G_0 D^\dagger D -G_1
\vec{\pi}^\dagger\hspace{-2pt} \cdot
\hspace{-2pt}\vec{\pi} - \chi\vec {\cal
Q}\hspace{-2pt} \cdot \hspace{-2pt}\vec {\cal Q} + \kappa\vec S
\hspace{-2pt}\cdot\hspace{-2pt} \vec S
\label{V}
\end{eqnarray}
where $G_0$, $G_1$, $\chi$ and
$\kappa$ are the interaction strengths of $d$-wave singlet
pairing, $d$-wave triplet pairing, staggered magnetization,
and spin--spin interactions, respectively.
Since $\ev{C_{su(4)}}$ is a constant,
by using Eq.\ (\ref{csu4})  we can eliminate
the $\vec \pi^\dagger \cdot
\vec \pi$ term and
after renormalizing the interaction strengths
the $SU(4)$ Hamiltonian
can be expressed as 
\begin{eqnarray}
H &=& H'_0-G [\ (1-p)D^\dagger D + p\vec {\cal Q}\hspace{-2pt} \cdot
\hspace{-2pt}\vec {\cal Q}\ ] + \kappa\tsub{eff}\, \vec S
\hspace{-2pt}\cdot\hspace{-2pt} \vec S
\label{generalH}
\label{eq20}
\\
H'_0 &=& \varepsilon\tsub{eff}\, n + \tfrac12 \mbox{v}\tsub{eff}\, n(n-1)
\end{eqnarray} 
with  $(1-p)G=G^{0}\tsub{eff}$, $pG=\chi\tsub{eff}$, and
$\kappa\tsub{eff}$ 
as the effective interaction strengths, and where $0\leq p \leq
1$ for the parameter $p$.
Since in this paper we primarily address
the ground state properties where $S = 0$,
the last term in Eq.\ (\ref{generalH}) will 
generally not enter into the later discussion. 

\section{The Dynamical Symmetry Limits}

As we have already noted, there are three subgroup chains of the $SU(4)$
symmetry that conserve spin and charge.  These define three dynamical
symmetries with clear physical meanings.  
The three dynamical symmetry limits $SU(2)$, $SO(4)$, and $SO(5)$, correspond to the
choices $p=0$, 1, and 1/2, respectively, in Eq.\ (\ref{eq20}).
The Hamiltonian, eigenfunctions, energy spectrum and the corresponding quantum 
numbers of these symmetry limits are listed in Tables I and II,
where we introduce a doping parameter $x$ that is 
related to particle number and lattice degeneracy 
through 
\begin{equation}
x=1-\frac{n}{\Omega}. 
\label{holedoping}
\end{equation}
The pairing gap $\Delta$ (measure of pairing order)
and the staggered magnetization (measure of AF order)
$Q$,
\begin{equation}
\Delta=G^{0}\tsub{eff}
\langle D^\dagger
D\rangle^{1/2}
\qquad
Q= \langle\vec{\cal Q} \hspace{-2pt}\cdot\hspace{-2pt}
\vec{\cal Q}\rangle^{1/2},
\end{equation}
may be used to characterize the states in these symmetry limits.
As we shall now see, each 
limit represents a  
different possible low-energy phase of the 
$SU(4)$ system.

\subsection{The SO(4) Limit}

The dynamical symmetry chain
$$
 SU(4) \supset SO(4) \times U(1) \supset SU(2)\tsub{s} \times U(1), 
$$
which we shall term the $SO(4)$ limit, corresponds to long-range 
AF order.  
This is the symmetry limit of Eq.\ (\ref{eq20})
when $p=1$.  
The $SO(4)$ subgroup is locally isomorphic to $SU(2)_F\times SU(2)_G$,
where the product group is generated by the linear combinations
\begin{equation}
\vec F= \tfrac12 (\vec {\cal Q} + \vec S)
\qquad
\vec G = \tfrac12 (\vec {\cal Q}-\vec S)
\label{su2xsu2}
\end{equation}
of the original $SO(4)$ generators $\vec {\cal Q}$ and $\vec S$.
We may interpret the new generators $\vec F$ and $\vec G$ physically by noting
that if we
transform $Q_{ij}$ and $S_{ij}$ defined in Eq.\ (\ref{E1}) to the physical
coordinate lattice, 
\begin{equation}
\begin{array}{c}
\begin{displaystyle}
            Q_{ij}=\sum_r (-)^r  c^{\dag}_{ri} c_{rj}
                     =\sum_{r={\rm \scriptsize even}} c^{\dag}_{ri} c_{rj}
                      -\sum_{r={\rm \scriptsize odd}} c^{\dag}_{ri} c_{rj}
\label{Qcoord}
\end{displaystyle}
\\[15pt]
\begin{displaystyle}
              S_{ij}=\sum_r c^{\dag}_{ri} c_{rj}
                     =\sum_{r={\rm \scriptsize even}} c^{\dag}_{ri} c_{rj}
                      +\sum_{r={\rm \scriptsize odd}} c^{\dag}_{ri} c_{rj}.
\end{displaystyle}
\label{Scoord}
\end{array}
\end{equation} 
implying that 
$\vec F$ is the generator of total spin on even sites
and $\vec G$ is the generator of total spin on odd sites.  
Thus, we may interpret the $SO(4)$
group as being generated by two independent spin 
operators:  one that is the total spin on all sites and one that is the
difference in spins on even and odd sites of the spatial lattice.   
This clearly is an algebraic version of the physical picture
associated with AF long-range order.

The $SO(4)$ Casimir
operator may be expressed as 
\begin{equation}
C_{so(4)} =2(\vec{F}\, ^2 +\vec{G}\, ^2).
\end{equation}
The
$SO(4)$ representations can be labeled 
by the spin-like quantum numbers 
$( F=w/2, G=w/2)$ 
where $w=N-\mu$ with $\mu=0,2,\ldots,N$. The eigenstates are labeled by $w$ and
the spin $S$, $\psi(SO4)=|N,w,S,m_s\rangle$,
and are of dimension $(w+1)^2$. 

The ground state
corresponds to $\omega=N$ and $S=0$, and has 
$n/2$ spin-up electrons on the
even sites ($F=N/2$) 
and $n/2$ spin-down electrons on odd sites
($G=N/2$), or vice versa. 
Thus it has maximal staggered magnetization 
\begin{equation}
Q=\tfrac12 \Omega(1-x)= \tfrac12 n
\end{equation}
and a large energy gap
(associated with the 
correlation 
$\vec{\cal Q}\hspace{-2pt}\cdot\hspace{-2pt}\vec{\cal Q}$) 
\begin{equation}
\Delta E = 2\chi\tsub{eff}(1-x)\Omega
\end{equation}
that inhibits electronic excitation and favors
magnetic insulator
properties at half filling. 
In addition, the pairing gap 
\begin{equation}
\Delta =  
\tfrac12 G^{0}\tsub{eff} \Omega \sqrt{x(1-x)}
\end{equation}
 is small near half
filling ($x=0$).
We conclude that these $SO(4)$ states are identified naturally with an
AF insulating phase of the system.

\subsection{The SU(2) Limit}

The dynamical symmetry chain
$$
 SU(4) \supset SU(2)\tsub{p} \times SU(2)\tsub{s} \supset 
SU(2)\tsub{s} \times U(1),
$$
which we shall term the $SU(2)$ limit, corresponds to 
SC order and is the
$p=0$ symmetry limit of Eq.\ (\ref{eq20}). 
The eigenstates are labeled by $v$ and spin $S$, 
$\psi(SU2)=|N,v,S,m_s\rangle$,
and are of dimension $(v+1)(v+2)/2$. 
The seniority-like quantum number $v$ is the number of
particles that do not form singlet $d$ pairs (see Table II). 
The ground state 
has
$v=0$, implying that all electrons are singlet-paired. In addition, there
exists a large pairing gap 
\begin{equation}
\Delta E = G^{(0)}\tsub{eff} \Omega 
\end{equation}
(see Table II), the pairing correlation is the largest among the three
symmetry limits, and the staggered
magnetization vanishes in the ground state:
\begin{equation}
\Delta = \tfrac12 G^{0}\tsub{eff} \Omega \sqrt {1-x^2}\ ,
\qquad
Q=0
\label{DeltaQ}
\end{equation}
Thus we propose that this state is a
pair condensate associated with a $d$-wave SC  
phase of the cuprates.

\subsection{The SO(5) Limit}

The dynamical symmetry chain
$$
 SU(4) \supset SO(5) \supset SU(2)\tsub{s} \times U(1),
$$ 
which we shall term the $SO(5)$ limit,
 corresponds to a phase with the nature of a transitional or 
{\em critical dynamical symmetry}.
This symmetry limit appears when $p=1/2$
in Eq.\ (\ref{eq20}).
The $SO(5)$ irreps are labeled by a quantum number $\tau$ and 
the eigenstates may be labeled by 
$\tau$ and the spin $S$, $\psi(SO5)=|N,\tau,S,m_s\rangle$ with
$N=\Omega/2-\tau +\lambda$, where $\lambda$ is the number of $\pi$ pairs. 
The irreducible representation dimensionality for given $N$ is 
$(\lambda +1)( \lambda +2)/2$ and  
the ground state has $\lambda=0$ and $S=0$. 

The $SO(5)$ dynamical symmetry has very unusual behavior. Although the
expectation values of 
$\Delta$ and $Q$ for ground state in this symmetry limit
are the same as that of Eq.\ (\ref{DeltaQ}) for the $SU(2)$ case, there
exists a huge number of states with different values of
$\lambda$ (the number of $\pi$ pairs) that can mix easily 
with the ground state when $x$ is small because the excitation energy in
this symmetry limit is  
\begin{equation}
\Delta E=\lambda G^{(0)}\tsub{eff}\Omega_{} x
\end{equation}
(see Table
II).
In particular, at half filling ($x=0$) the ground state is highly degenerate
with respect to $\lambda$ and 
mixing different numbers of $\pi$ pairs in the ground state costs no
energy.  The $\pi$ pairs must be 
responsible for the antiferromagnetism in this phase, since 
within the model space only $\pi$ pairs carry spin. 
Thus the ground state 
in this symmetry limit has 
large-amplitude fluctuation in the AF order (and SC order). This 
indicates that the $SO(5)$ symmetry limit 
is associated with phases 
in which the system is extremely susceptible to fluctuations between 
AF and SC order.

\subsection{Energy Surfaces}

The soft nature of the $SO(5)$ phase 
is seen most clearly if we introduce approximate solutions
in terms of $SU(4)$
coherent states \cite{wmzha90}.  
We shall discuss the interpretation of $SO(5)$
using coherent states in a separate paper \cite{lawu99}, 
but we quote one result of
that study here.  
In Fig.\ 2, ground-state energy surfaces 
for various particle number $n$ (or doping $x$) are plotted
as a function of a quantity $\beta$, which is related directly to the
AF order parameter (see figure caption). Three plots are associated
with the symmetry limits discussed above $(p=0, 1, 1/2)$, and one corresponds to
a situation with a slight $SO(5)$ symmetry breaking ($p= 0.52$).
For all doping values $x$ the energy minimum lies at
$\beta = 0$ (implying that $Q$=0) in the $SU(2)$ limit (Fig.\ 2a), while 
it lies at 
$$
\beta = \sqrt{\tfrac14 (1-x)}
$$
(implying that $Q=n/2$) for the $SO(4)$ limit (Fig.\ 2b).
In Fig.\ 2c,
there is a broad range of doping in which the $SO(5)$ energy surface is
almost flat in the parameter $\beta$,
implying large excursions in the
AF (and SC) order:  one can fluctuate into the other at negligible energy cost.
Notice in Fig.\ 2d that as doping varies the
$SO(5)$ Hamiltonian with slight symmetry breaking
interpolates smoothly between AF order at half filling
and
SC order at smaller filling.
Thus $SO(5)$ acts as a kind of doorway between $SU(2)\tsub p$ and $SO(4)$
symmetries and
this gives a precise meaning to the
assertion \cite{zha97}
that the $SO(5)$ symmetry rotates between AF and
SC order.

However, the present discussion shows that the relation between the
AF and SC 
phases is more complex than a simple $SO(5)$ rotation because of the non-abelian
nature of the $SU(4)$ parent group of $SO(5)$.
Such  dynamical symmetries
that
define a phase of the theory but that also interpolate between two other
dynamical
symmetries are known in nuclear structure physics where they have been
termed  {\em critical dynamical symmetries} \cite{wmzha88}.  

The soft $SO(5)$ energy surface could lead to ``spin-glass-like'' phases
(by which we mean phases with local AF or SC order but with large fluctuations
in both).  It could
also lead to inhomogenous
 structure like stripes if there is a periodic spatial modulation of the
system, since the soft nature of the energy surface implies that relatively small
perturbations can shift an $SO(5)$ system between AF and SC  
behavior.  As noted in an earlier footnote, perturbations of the symmetry by 
a charge density wave could be one source of such a spatial modulation.

\section{Phases and Phase Transitions}

In quantitative tests of the $SU(4)$ model with parameters determined by
fitting to pairing gap and pseudogap data, a cuprate phase diagram has been
predicted.  It is found that for the symmetry mixing parameter $p$ close to but 
larger than $0.5$  (antiferromagnetically perturbed $SO(5)$ symmetry) 
the experimental data are described quite well.
The results of this study will be published separately \cite{sun99,wu00}.
In this paper, we concentrate on general features and 
use the preceding discussion to
construct the qualitative phase diagram
illustrated in Fig.\ 3.  
This diagram, which contains essential features of the realistic
phase diagram, is constructed based on the following arguments.

\subsection{Phase Diagram}

Consider the Hamiltonian (\ref{generalH}) and, to simplify this discussion,
let us assume
the spin term $k\tsub{eff}\vec{S}\cdot\vec{S}$ to be neglected.
In that case, there are two fundamental energy scales in the
Hamiltonian (\ref{generalH}): $H'_0$ and $ H - H'_0 \sim G\Omega^2$.  
The term
$H'_0$ originates in the single-particle energies
$\varepsilon\tsub{eff} n$ 
and the $SU(4)$ invariant $C_{su4}$.  This term depends only 
on particle number and is isotropic:
it has the same expectation value for any state
in the $SU(4)$ representation space.  In contrast, the other terms  $H - H'_0$
(with characteristic
energy scale $G\Omega^2$) are anisotropic in the $SU(4)$ space:
states associated with different dynamical symmetries may have quite different
expectation values.  (Thus, if we make a mean-field approximation to the
present many-body theory---like the method of 
coherent states---these anisotropic
terms will lead to spontaneously broken
symmetries.)

The term 
$H'_0$ in the Hamiltonian (\ref{generalH})
 may be regarded as the energy scale for the
$U(4)\supset U(1)\times SU(4)$ symmetry,  which 
describes 
a fermion system in which
electrons are all paired but with no distinction among the $d$ pairs and
$\pi$ pairs. 
We may
expect this symmetry to hold while the thermal energy is  
less than $H'_0$. 
When the system is cooled, the thermal energy eventually drops below
the anisotropic scale $G\Omega^2$, the anisotropic
pairing and antiferromagnetic correlations $H - H'_0$ 
become important, and
$SU(4)$ breaks to its subgroups. 
Then different low-temperature phases appear,
with the favored phases  
depending on the competition
between $D^\dagger D$ and $\vec
{\cal Q}\cdot \vec {\cal Q}$ 
interactions as a function of  
doping concentrations.  

From the preceding discussion, at zero temperature we
expect the $SO(4)$
AF phase to dominate at half filling if $p>0.5$, because
the energy surface is minimized at $\beta=[\tfrac14 (1-x)]^{1/2}$, implying
that the staggered magnetization 
is large:  $Q= n/2$.  On the other hand, at
larger doping the $SU(2)$ pairing phase is favored 
(the pairing energy is optimized and the staggered magnetization
minimized) (see Fig.\ 2d).  
Finally, the intermediate doping region is described naturally by
the $SO(5)$ critical dynamical symmetry
 that interpolates between SC and AF behavior with
doping.  Thus, $SU(4)$ symmetry implies the schematic phase diagram of Fig.\ 3,
independent of detailed calculations.

\subsection{Phase Transitions and Symmetry Breaking}

In the Hamiltonian (\ref{eq20}), the parameter $p$ takes on the values 0, 1,
and 1/2 in the three symmetry limits of the theory. For any other allowed
value of $p$ the system exhibits $SU(4)$ symmetry but there is no dynamical
symmetry (If $p \ne 0, 1, 1/2$,
the eigenstates of the system are linear superpositions of
eigenstates from the three dynamical symmetry chains).
In this case, phase transitions are driven by
microscopically-determined control parameters that change the
expectation value of the terms of the Hamiltonian such that $\langle\vec {\cal 
Q}\hspace{-2pt}\cdot\hspace{-2pt}\vec{\cal
Q}\rangle$ is dominant in some circumstances while $\langle
D^\dagger\hspace{-2pt}\cdot\hspace{-2pt}D\rangle$ dominates in others.
This permits phase transitions among the
AF ($SO(4)$ limit), $SO(5)$, and SC phases ($SU(2)$) to be studied using a 
fixed Hamiltonian. In cuprates, the hole-doping
$x$ is an example of such a microscopically-determined control parameter. 
Thus, a Hamiltonian that possesses an approximate $SO(5)$
symmetry (antiferromagnetic perturbed $SO(5)$) can 
have AF solutions at small hole-doping and SC
solutions at large hole-doping. We shall give an explicit example of a phase 
transition driven microscopically by changing hole doping in the next section.

\section{The Transition between Antiferromagnetism and Superconductivity}

Above, we have used  the method of generalized coherent states to
give an
interpretation of the $SO(5)$ subgroup as a critical dynamical
symmetry interpolating between AF  
 and SC order and having the character of a
``spin glass'' or perhaps leading to stripe phases for a large
range of intermediate doping parameters
\cite{lawu99}.  In this section we address further the relationship between the
other two
phases of the theory.  We show that the microscopic symmetry incorporates a
differing
dependence on doping for SC and AF order.  This implies that the group
structure itself
controls the transition between the superconducting $SU(2)$ symmetry and the 
antiferromagnetically ordered $SO(4)$ symmetry.  Thus, we conclude that the
$SU(4)$
symmetry leads naturally to AF 
 order at half-filling and to $d$-wave
superconductivity   
as the system is hole-doped away from half-filling for a broad range of
Hamiltonian parameters.

\subsection{A Simplified Picture}

To simplify the discussion, we drop the common dependence of both phases on the
spin and
charge generators and consider the competition between the $SO(4)$
stabilization energy
coming from a term $\vec {\cal Q} \cdot \vec {\cal Q}$
and the $SU(2)$ stabilization energy coming from a term $D^\dagger D$ in the
Hamiltonian.  
These clearly have fundamentally different behaviors with changing particle
number.  From Table II and Eq.\ (\ref{holedoping}), we may conclude that 
in the respective ground states,
\begin{equation}
\begin{array}{c}
\ev{\chi \vec {\cal Q} \cdot \vec {\cal Q}} = \chi N(N+2)
\\[5pt]
\ev{G_0 D^\dagger D} =G_0  N(\Omega -N+1)
\label{compare}
\end{array}
\end{equation}
where the pair number is $N=\tfrac12 n$.
At half filling ($N=\Omega/2$), if $\chi /G_0>1$, the Hamiltonian 
exhibits an effective $SO(4)$ symmetry because the $SO(4)$ correlation energy
$\vec {\chi \cal Q}\cdot\vec {\cal Q}$ dominates. As the
particle number decreases (hole-doping $x$ increases), the $SO(4)$
correlation energy decreases quadratically but the $d$-wave
pairing $D^\dagger D$ decreases much more slowly (essentially linearly). 
Therefore, the pairing correlation $D^\dagger D$ 
will eventually dominate and the Hamiltonian exhibits effective $SU(2)$ symmetry.

These features imply
immediately that, if $\chi /G_0>1$,
antiferromagnetism will tend to dominate at
half-filling but,
as the system is 
doped away from half-filling with holes, eventually pairing will dominate
and the system will become superconducting.  The transition
point  will depend on the relative strengths of the 
$D^\dagger D$ and $\vec {\cal Q} \cdot \vec {\cal Q}$ terms in the effective
Hamiltonian, but the AF ground state at half filling and the
tendency to superconductivity as the system is doped away from 
half-filling is a direct consequence of the group structure,
independent of detailed parameter choices.

\subsection{Analogies in Nuclear Structure}

This competition between antiferromagnetism and superconductivity bears many
strong resemblances to the competition in nuclear physics between spherical
and deformed structure for nuclei.  There, it has been shown that the transition 
from spherical nuclei,  which dominate
the beginnings and endings of shells, to deformed nuclei, which often dominate the
middle of shells, is controlled by the microscopic competition between long-range
quadrupole-quadrupole interactions favoring deformation 
and short-range monopole pairing interactions
favoring spherical vibrational structure.  This competition 
in nuclear structure may be 
expressed algebraically as a competition between a dynamical symmetry that
favors  pairing and a dynamical symmetry that favors multipole (particle--hole)
interactions
\cite{wu94}.  

The essential physics of the spherical--deformed transition in nuclear
structure physics has
been shown to be determined by the differing particle number dependence of the
dynamical
symmetries:  nuclear pairing energy increases linearly with particle number but
the quadrupole
deformation energy is quadratic in particle number (that is, essentially the
same relationship as for Eq.\ (\ref{compare})).  Thus, the group structure
dictates that  
spherical vibrational nuclei (favored by pairing energy)
dominate the ends of shells and deformed nuclei (favored by the deformation
energy)
dominate the
middle of shells
\cite{footnote2}.  
This behavior is a close analog of the competition between
antiferromagnetism dominating the half-filled lattice and superconductivity
dominating the
hole-doped lattice that we have discussed in this paper.  

This analogy between condensed matter and nuclear
physics might not be accidental.  The structure of heavy nuclei and of
high-temperature
superconductors both correspond to complex many-fermion problems in which 
strong
particle--hole and pairing interactions
involving $d$ pairs 
play pivotal roles.  It has been demonstrated 
that dynamical symmetries in the microscopic fermion degrees of freedom
provide a simple but powerful unifying principle for 
the nuclear structure problem \cite{wu94}.  Recent results, such as those
presented here and in Ref.\ \cite{zha97,NE}, suggest that related
symmetries may play a
similar role for condensed matter and that 
these problems from rather different fields of physics may have a common
dynamical
algebraic structure.

\section{SU(4) or SO(5)?}

Let us now discuss more precisely the
relationship between our $SU(4)$ model and the Zhang's $SO(5)$ model 
\cite{zha97}.  Although the methodologies used to obtain the two models are
rather different, the Zhang $SO(5)$ group is a subgroup of our
$SU(4)$ group and both model Hamiltonians 
possess antiferromagnetic perturbed SO(5) symmetry, 
implying that the two models are closely related to each other.
Our $SU(4)$ model and Zhang's $SO(5)$ model 
have the same building blocks (the operator set (\ref{E4}), 
but see note \cite{u1note}).
The essential
difference is that we implement the full quantum dynamics
(commutator algebra) of these
operators exactly, while in Ref.\ \cite{zha97}, the dynamics is
implemented in an approximate manner:
a subset of 10 of the operators
acts as a rotation on the remaining 5 operators
$\{D^\dagger,D,{\cal \vec{Q}}\}$, which are treated phenomenologically
as 5 independent
components of
an order-parameter vector (superspin
$\vec{n}$\,).  Thus only 10 of the 15 generators of our $SU(4)$ are 
treated fully dynamically in the Zhang $SO(5)$.  If the full quantum
dynamics (full
commutator algebra) of the 15 operators is taken into account, the symmetry is
$SU(4)$, not $SO(5)$.

The embedding of $SO(5)$ in our larger algebra has various physical
consequences that do not appear if the $SO(5)$ subgroup is considered in
isolation.
We list four:

1. As we have seen, a phase transition  from antiferromagnetism to
superconductivity 
at zero temperature that is controlled by the
doping emerges naturally and microscopically,
whereas in the $SO(5)$ model a symmetry-breaking term proportional to
a chemical potential
has to be introduced by hand.  This occurs because our $SO(5)$ is embedded in a
larger group that contains generators breaking $SO(5)$ but preserving $SU(4)$.

2. The present results suggest that the $SO(5)$ subgroup
is only one of the symmetries relevant to the cuprate problem.
It is a transitional symmetry that links AF to 
SC behavior,
suggesting that it is most useful for the underdoped region. 
But the AF phases at half filling and the optimally
doped superconductors
are more economically described by our $SO(4)$ and $SU(2)\tsub p$
symmetries, respectively.  All are unified in the $SU(4)$ highest symmetry.

3. As we discuss in a separate publication \cite{sun99},
the present $SU(4)$ theory leads naturally to the appearance of pseudogap
behavior \cite{TS99}, which occurs above the  
SC transition temperature $T_c$. 
The embedding of $SO(5)$ in $SU(4)$ is central to this behavior. 

4. The different methodology of the $SU(4)$ dynamical symmetry approach
suggests a different physical interpretation of the  $SO(5)$ subgroup symmetry.
We suggest that it should be viewed as an effective symmetry 
that operates in a severely
truncated space.  As we shall elaborate below,
this implies that its microscopic validity is a question of
the physical correctness of the matrix elements evaluated in that truncated
model space, not
of whether a particular Hamiltonian with some relevance for the full space
(Hubbard, for example) possesses such a
symmetry. 

Thus, both the Zhang $SO(5)$ theory and the present $SU(4)$ theory imply the
existence of an approximate $SO(5)$ symmetry in the Hamiltonian of high-temperature
superconductors.  However, the $SU(4)$ theory  encompasses a broader range of
physics that may be relevant for cuprate superconductor.  
Further, as we have argued, a unified quantum mechanical
treatment of the generators and order parameter vector of the Zhang $SO(5)$   
implies that the full quantum mechanical
symmetry is $SU(4)$ and not $SO(5)$.

\section{Discussion}

The $U(4) \supset U(1) \times SU(4)$ symmetry
represents a fully microscopic fermion system in which SC 
and AF modes 
enter on an equal
footing.  At this ``unification'' level, there is in a sense no distinction
among these degrees of freedom, just as in the Standard Electroweak Theory
of elementary particle physics
the electromagnetic and weak interactions are unified above the intermediate
vector boson mass scale. 
We may expect this symmetry to hold while the
temperature of the system is sufficiently high that anisotropic quantum
fluctuations in the directions associated with these collective degrees of
freedom can be neglected (recall that the $SU(4)$ scale is set by $H'_0$ and
the anisotropic scale by 
$G\Omega^2$), but not so high that thermal fluctuations destroy the
integrity of the collective modes corresponding to the $SU(4)$ symmetry.
This $SU(4)$ region then corresponds to the pseudogap regime \cite{sun99}.  

Generalizing
language already introduced by Zhang \cite{zha97}, in this regime we may view
the system as having condensed into $SU(4)$ pairs, which fixes the length of
the state vectors (through the $SU(4)$ Casimir expectation value) but not their
direction in the state-vector space.  Physically, this means that the system is
paired, with the pair structure exhibiting $SU(4)$ symmetry, but is neither
SC nor AF because fluctuations in those directions
in the $SU(4)$ space are small relative to the scale set by the temperature.
Stated in another way, the $SU(4)$ pairs are of collective strength, but are
not condensed into a state with long-range order.  Stated in yet another way,
neither the AF nor SC order parameters have finite
expectation values in this regime, but a sum of their squares (the $SU(4)$
Casimir) does.  This constraint implies an intimate connection between 
superconductivity and antiferromagnetism 
in the $SU(4)$ model.  They are, in a sense, different projections of the same
fundamental vector in an abstract algebraic space.

Compared to the Hubbard or {\em t-J} models, the dynamical symmetry approach 
applied here is just a different way to simplify 
a strongly-correlated electron system. 
In the Hubbard or {\em t-J} models, approximations
are made to simplify the Hamiltonian but no specific truncation is assumed for
the configuration space, although 
in  practical calculations a truncation is typically 
necessary.  In contrast, we make no
approximation to the Hamiltonian.  
The only approximation is the (severe) space truncation. 
The symmetry dictated Hamiltonian includes all possible interactions in the
truncated space.    
In principle, if all the degrees of freedom of the system are included, this
approach constitutes  an exact microscopic theory.
The validity of this approach depends
entirely on the validity of the
choice of truncated space and its effective interactions, 
which may be tested by comparison with the data.

Thus, we 
suggest that effective low-energy theories of the type
discussed here need not have much direct relation to the Hamiltonian or
wavefunctions of the Hubbard or {\em t-J} models,  because both the Hamiltonian and
the wavefunctions in different model spaces could be very different.
What is observable quantum mechanically is
the matrix elements, not operators or wavefunctions separately.   As the
$SU(4)$ theory makes clear, we may view these dynamical
symmetries as operating in a severely truncated collective subspace 
in which the truncation
has been implemented primarily by symmetry considerations and only secondarily by
energy criteria.  Thus, it is possible that the matrix elements of the $SU(4)$
theory and a Hubbard or {\em t-J} 
model calculation might be comparable, even if the 
Hamiltonians and wavefunctions are separately
quite different.  

The advantage of the dynamical symmetry approach is its cleanness and
simplicity.  It is clean because the {\em only} approximation is the 
selection 
of the truncated model space. Thus, a failure
of the method is a strong signal
that one has chosen a poor model space.
It is
simple because the method supplies analytical solutions for various dynamical 
symmetry
limits as a starting point.  These symmetry-limit solutions provide an 
immediate
handle on the physics and permit an initial judgement of the model's validity
without large-scale numerical calculations.  Beyond the
symmetry limits, numerical calculations are necessary.  However, because of 
the
low dimensionality of the models spaces and the power of group theory, such
numerical calculations are much simpler
than those in, say, a Hubbard or {\em t-J} model.

The primary limitation of the dynamical symmetry approach is that the 
interactions
of the model space are necessarily effective interactions that will generally 
be
strongly renormalized by the symmetry-constrained space truncation.  Thus,
their strengths must be supplied separately from the dynamical symmetry method 
itself, either
phenomenologically (by fitting model-space interaction strengths to data), or 
by
a microscopic study of the relationship between the effective and full-space 
Hamiltonians.

Finally, let us counter a possible philosophical criticism of the approach
taken here.  
Simple symmetries as a predictor of dynamics has
found powerful application in fields such
as particle physics or nuclear physics. 
However, there is a common point of view 
that the possible ground states in 
high-temperature superconductors are too complex 
to
permit a simple model like the current one (which operates in a drastically
truncated subspace having simple wavefunctions and highly stylized operators)
to have any validity.  
Although one can make such an argument formally, this ignores 
the rather obvious point that nature has managed to construct a stable
ground state having well-defined, collective  
properties that change systematically
from compound to compound.  Extensive
experience in many fields of physics
suggests that this is the 
signal that the phenomenon in question is described
by a {\em small effective subspace with renormalized interactions} (that may
differ substantially in form and strength from those of the full space) and
governed by a symmetry structure of manageable dimensionality.
Thus, if an approach like the one proposed here gives
correct results for highly non-trivial phenomenology like the doping dependence
of observable quantities, one must take seriously the possibility that the
corresponding small symmetry-dictated
subspace may have relevance to the effective behavior
of real physical systems, no matter how complex they may appear to
be superficially.

\section{Summary and Conclusions}

In summary, 
an $SU(4)$ model of High-$T_c$ superconductivity has been  proposed 
that contains three dynamical symmetries:
A SC phase identified with the
$SU(2)\tsub{p}$
dynamical symmetry, an AF phase  identified with the
$SO(4)$ dynamical symmetry, and an $SO(5)$
phase extremely soft
against AF and SC 
fluctuations over a substantial doping fraction
that serves as a critical dynamical symmetry interpolating between the other two
phases. 
Realistic systems may mix these sub-symmetries while 
retaining an approximate $SU(4)$ symmetry.  Zero-temperature
phase transitions 
are shown to be driven 
by the competition between the $d$-wave pairing and
the AF $\vec{\cal
Q}\cdot\vec{\cal Q}$ interactions, as controlled 
microscopically by
the hole-doping concentration.  
This model leads naturally to the appearance of 
pseudogaps in the underdoped regime because it introduces multiple energy
scales that permit pairs to form before they condense into states with
long-range order.

Thus, we propose that high $T_c$ behavior of the cuprates results from  an
$SU(4)$ symmetry realized dynamically, and because this symmetry is microscopic
its physical interpretation is accessible to calculation.  This 
provides a solvable model incorporating many features of cuprate
superconductors, a possible understanding of the cuprate phase 
diagram as arising from dynamically-realized symmetries, and substantial
insight concerning recent $SO(5)$ theories of $d$-wave superconductivity.

\acknowledgments

We thank Eugene Demler, Shou-Cheng Zhang,
Pengcheng Dai, and Ting-Guo Lee for useful discussions.
L. A. Wu was supported in part by the National Natural Science Foundation of
China.  C.-L. Wu is supported by the National Science Council of ROC.

\baselineskip = 14pt
\bibliographystyle{unsrt}

\begin{thebibliography} {99}
\bibitem{minimalnote}
One can consider even larger algebras, which have the advantage of
incorporating a broader range of physics but at the expense of additional
parameters.  We adopt the point of view that one should first investigate
thoroughly the minimal algebra consistent with the degrees of freedom that are
likely to be important for the problem at hand.
\bibitem{spontaneousnote}
The reader should recall that spontaneous symmetry
breaking is a useful concept but that the name is 
a misnomer.  A spontaneously broken symmetry corresponds in reality
to a conserved symmetry that has been hidden but is still present; it merely
manifests itself in a different way.  The physical states for a finite system
must be constructed by restoring any 
symmetries broken by approximate solutions, for example through projection
techniques.  Thus, BCS solutions are in many respects an excellent
description of superconductors, but they are approximate and violate particle
number conservation.  The true physical states of a superconductor must then
correspond to a projected combination of BCS states that conserves particle
number.  Of course, as a practical matter, restoration of the symmetry may not
be too important if the effect of the symmetry violation is small for the
physical quantity of interest.
\bibitem{cdwavenote}
If only part of the system is assumed to constitute the
$d$-wave subspace (for example, in a 2-fluid model), the $U(1)$
factor can produce charge density wave excitations in the
non-superfluid space.  Because of the direct product structure,
in the symmetry limits these charge density waves would be
independent of any excitations in the $d$-wave subspace. In the
presence of symmetry breaking, the charge density waves could
perturb the $d$-wave subspace and generate inhomogenous structures
like stripe phases, as we discuss in section IV.D.
\bibitem{threebodynote}
Groups generally may have more than one Casimir invariant.  We shall use the
term ``Casimir'' to refer loosely to the 
lowest-order such invariants (which are generally quadratic in the
group generators).  In the context of the present discussion,
quadratic Casimirs are associated with 2-body interactions at the microscopic
level.  Higher-order Casimirs are then generally associated with 3-body and
higher interactions.  The restriction of our Hamiltonians to polynomials of
order 2 in the Casimirs is then a physical
restriction to consideration of only 1-body and  
2-body interactions.
\bibitem{u1note}
In addition, 
there is the $U(1)$ generator of charge
density waves in our full $U(4) \supset U(1) \times SU(4)$ algebra that does
not appear in the Zhang $SO(5)$ algebra.
\end{thebibliography}

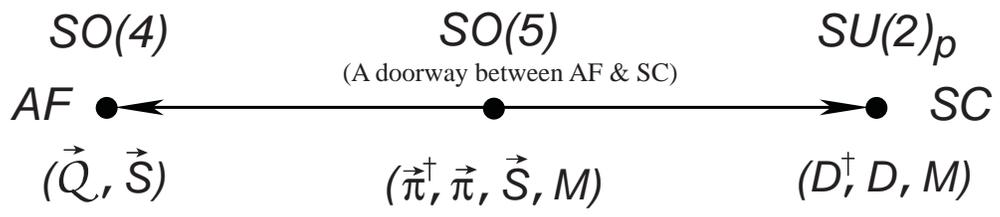
\begin{figure}
\caption{Dynamical symmetries associated with the $U(4)$ symmetry.  
The generators are listed for each subgroup.}
\label{fig1}
\end{figure}

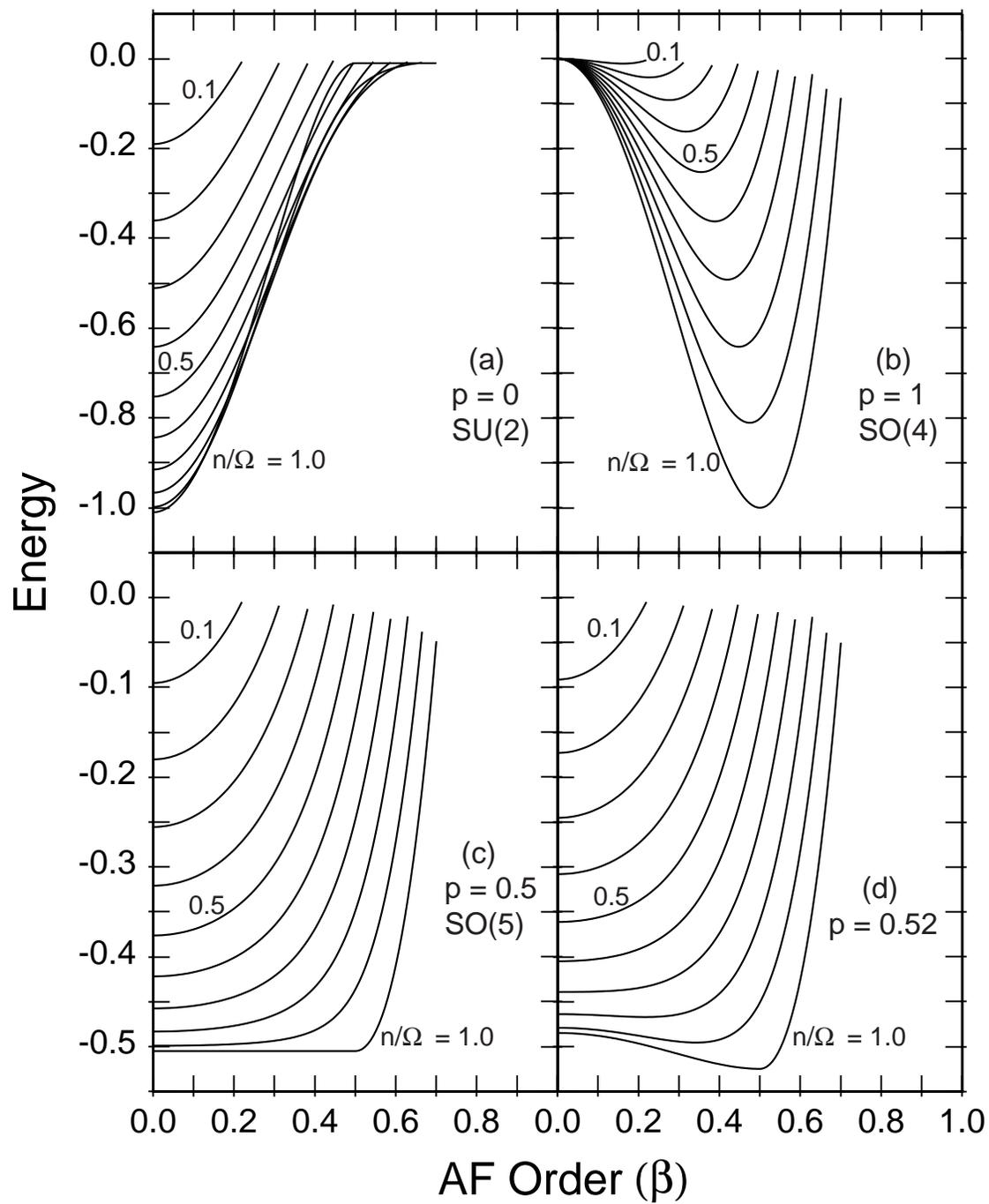
\begin{figure}
\caption{Coherent state energy surfaces for
Eq.\ (\protect\ref{generalH}).
The energy unit is
$G\Omega^2 / 4$ (see Eq.\ (\protect\ref{eq20})).
The staggered magnetization
$Q$ is related to $\beta$ by
$\langle Q \rangle = 2\Omega \beta_0 (n/2\Omega -\beta_0^2)^{1/2}$,
where $\beta_0$ is the value
of $\beta$ at the stable point, so
$\beta$ measures the AF order.
Numbers on curves are the lattice occupation fractions,
with $n / \Omega = 1$ corresponding to half filling and $0 < n / \Omega < 1$
to finite hole doping.
}
\label{fig2}
\end{figure}

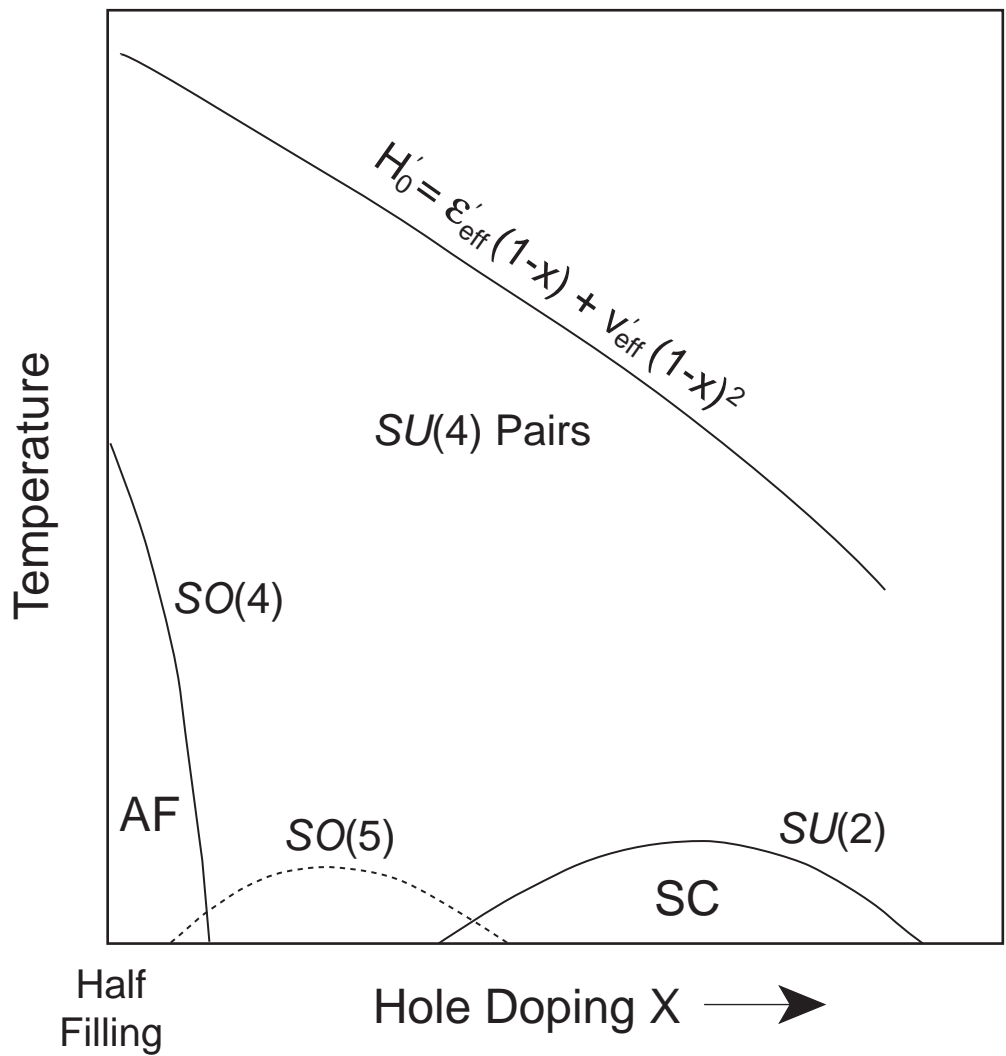
\begin{figure}
\caption{Schematic phase diagram for $SU(4)$ symmetry based on Fig.\ 1d.
The $H'_0$ in Eq.\ (\protect\ref{generalH}) 
is expressed in terms of hole doping $x$ with
$x=1-n/\Omega$;
$\varepsilon'_{\mbox{\protect\scriptsize{eff}}}=
\varepsilon_{\mbox{\protect\scriptsize{eff}}}\Omega/2$, and
v$'_{\mbox{\protect\scriptsize{eff}}}$ 
=v$_{\mbox{\protect\scriptsize{eff}}}\Omega^2/2$. }
\label{fig3}
\end{figure}

\newpage
\onecolumn

\begin{table}
\caption {Properties of $SU(4)$ and its subgroups.$^*$}
\vspace{4pt} 
\begin{tabular}{lcccc} 
Group & Generators& Quantum Numbers& Casimir
Operator& Casimir Eigenvalue
     \\   \tableline  \\  
$SU(4)$&
$Q_+$, $\vec S$, $\vec {\cal Q}$, $\vec \pi^\dagger$&
$\sigma_1=\tfrac{\Omega}{2}$ ($\sigma_2=\sigma_3=0$) &
$\vec \pi^\dagger \hspace{-2pt}\cdot \hspace{-2pt}\vec \pi
    + D^\dagger \hspace{-2pt} D +
    \vec S \hspace{-2pt}\cdot\hspace{-2pt} \vec S $&
$\tfrac{\Omega}{2}(\tfrac{\Omega}{2}+4)$ 
     \\ \vspace*{6pt}
 &\ $\vec \pi$, $D^\dagger$, $D$, $M$& &
$+ \vec {\cal Q}
    \hspace{-2pt}\cdot\hspace{-2pt} \vec {\cal Q} + M(M-4)$&  
     \\ 
$SO(4)$&
$\vec {\cal Q}$, $\vec S$ & $w$ ($F=G=w/2$)&
$\vec {\cal Q}
    \hspace{-2pt}\cdot\hspace{-2pt} \vec {\cal Q} + \vec S
\hspace{-2pt}\cdot\hspace{-2pt}
    \vec S$& 
$w(w+2)$  \vspace{6pt}
\\ 
$SO(5)$ &
$\vec S$, $\vec \pi^\dagger$, $\vec \pi$, $M$ &
$\tau$ ($\omega=0$) &
$\vec \pi^\dagger \hspace{-2pt}\cdot \hspace{-2pt}\vec \pi
          + \vec S \hspace{-2pt}\cdot\hspace{-2pt} \vec S+M(M-3)$ &
$\tau(\tau+3)$ \vspace{6pt}
     \\ 
$SU(2)\tsub p$ &
$D^\dagger$, $D$, $M$ &
$v$ &
$D^\dagger D + M(M-1)$ &
\vspace*{6pt} 
$\tfrac14 (\Omega-v)(\Omega -v +2)$ 
     \\ 
$SU(2)\tsub s$ &
$\vec S $ &
$S$ &
$\vec S \hspace{-2pt}\cdot\hspace{-2pt} \vec S$ &
$S(S+1)$
\end{tabular}
*{\footnotesize  Valid for representations with no broken pairs.}
\label{table1}
\end{table}

\begin{table}
\caption { \small \hspace{12pt}The Hamiltonian, eigenstates and spectra in three
dynamical symmetry limits of the $SU(4)$ model.$^\dagger$ \hspace{24pt}
$ E_{g.s.}$ is the ground state energy,
$\Delta E$ the excitation energy, $Dim$ the dimension of each representation for
a given
$N$,  $N = n/2$ the pair number, $x = 1-n/\Omega$, , and $\kappa\tsub{so4}=
\kappa_{\mbox{\scriptsize{eff}}}+\chi_{\mbox{\scriptsize{eff}}}$. } \vspace{4pt}
\begin{tabular}{c} {\scriptsize
$
\begin{array}{lll}\\
\vspace{8pt}
 \mbox{\bf SU(2) limit: } |\psi(SU2)\rangle=\ket{N,v,S,m_s}\ \ &\mbox{\bf SO(4)
limit: } |\psi(SO4)\rangle=\ket{N,w,S,m_S} \ \ &\mbox{\bf SO(5) limit:
}|\psi(SO5)\rangle=\ket{\tau,N,S,m_S}\\
\vspace{8pt}
 \ev{C_{SU(2)_p}}=\tfrac14(\Omega-v)(\Omega-v+2) &\ev {C_{SO(4)}}=w(w+2),\quad
w=N-\mu &\ev{C_{SO(5)}}=\tau(\tau+3),\quad \tau=\Omega/2-N+\lambda \\ 
\vspace{8pt} Dim(v,N)=(v+1)(v+2)/2 &Dim(w,N)=(w+1)^2
&Dim(\tau,N)=(\lambda+1)(\lambda+2)/2\\ 
\vspace{8pt}
 H =H'_0-G^{(0)}\tsub{eff} D^{\dagger}D+\kappa\tsub{eff}\ \vec
S\hspace{-2pt}\cdot\hspace{-2pt}\vec S  &H =H'_0-\chi_{\tsub{eff}}\vec {\cal Q}
    \hspace{-2pt}\cdot\hspace{-2pt} \vec {\cal Q}+\kappa\tsub{eff}\ \vec
S\hspace{-2pt}\cdot\hspace{-2pt}\vec S
 &H =H'_0-G^{(0)}\tsub{eff}\vec {\cal Q}
    \hspace{-2pt}\cdot\hspace{-2pt} \vec {\cal Q}+D^{\dagger}D
+\kappa\tsub{eff}\ \vec S\hspace{-2pt}\cdot\hspace{-2pt}\vec S\\
\vspace{8pt}
\ev{D^{\dagger}D} =\ev{
C_{SU(2)_p}-M(M-1)} &\ev{\vec {\cal Q}
    \hspace{-2pt}\cdot\hspace{-2pt} \vec {\cal Q}}=
\ev{C_{SO(4)}- \vec S\cdot\hspace{-2pt}\vec S} &\ev{\vec {\cal Q}
    \hspace{-2pt}\cdot\hspace{-2pt} \vec{\cal Q}+D^{\dagger}D}
=\ev{C_{SU(4)}+M-C_{SO(5)}}\\
\vspace{8pt}
 E_{g.s.}=H'_0-\frac{1}{4}G^{(0)}\tsub{eff}\Omega^2(1-x^2)
&E_{g.s.}=H'_0-\frac{1}{4}\chi\tsub{eff}\Omega^2(1-x)^2
&E_{g.s.}=H'_0-\frac{1}{4}\chi\tsub{eff}\Omega^2(1-x)^2\\
\vspace{8pt}
 \Delta E=\nu G^{(0)}\tsub{eff}\Omega+\kappa\tsub{eff}\ S(S+1),\hspace{2pt}
\nu=v/2 &\Delta E=\mu \chi_{\tsub{eff}}(1-x)\Omega+\kappa\tsub{so4}\ S(S+1)
&\Delta E=\lambda G^{(0)}\tsub{eff}\ x \Omega+\kappa\tsub{eff}\ S(S+1)\\
 \nu =0,1,2,\ldots N;\hspace{5pt} S=\nu, \nu-2, \ldots 0
\hspace{28pt}
 &\mu=0,2,\ldots N;\hspace{5pt} S=w, w-1, \ldots 0 
\hspace{42pt} &\lambda=0,1,2,\ldots, N ;\hspace{5pt} S=\lambda,
\lambda-2, \ldots 0
\nonumber
\end{array}
$}
\end{tabular}
$^\dagger${\footnotesize  Assume $N$ is even and $1/\Omega$ is negligible.}
\label{table2}
\end{table}

\end{document}